\pdfoutput=1

\documentclass[sigconf,screen]{acmart}
\AtBeginDocument{%
  \providecommand\BibTeX{{%
    \normalfont B\kern-0.5em{\scshape i\kern-0.25em b}\kern-0.8em\TeX}}}

\setcopyright{acmlicensed}
\acmDOI{10.1145/3663529.3663839}
\acmYear{2024}
\copyrightyear{2024}
\acmSubmissionID{fsecomp24industry-p37-p}
\acmISBN{979-8-4007-0658-5/24/07}
\acmConference[FSE Companion '25]{Companion Proceedings of the 33rd ACM International Conference on the Foundations of Software Engineering}{23 -- 27, 2025}{Trondheim, Norway}
\acmBooktitle{Companion Proceedings of the 33rd ACM International Conference on the Foundations of Software Engineering (FSE Companion '25), 23 -- 27, 2025, Trondheim, Norway}

\usepackage{tabularx}
\newcommand{\InitialDate}{{28th October 2024}}
\newcommand{\FinalDate}{{31st December 2024}}
\newcommand{\toolname} {{\sc ACH}}
\newcommand{\toolExpendedName} {{\sc Automated Compliance Hardener}}
\newcommand{\PrivacyMutants} {simulated privacy faults}

\newcommand{\ProductList} {Aloha, Facebook Feed, Instagram, Messenger, Oculus, Wearables, and WhatsApp}
\newcommand{\NumberOfProducts} {7}

\newcommand{\OverallAcceptRate} {73\%}
\newcommand{\PercentRelevant} {36\%}
\newcommand{\backtick}{{\char96}}
\newcommand{\triplebacktick} {{\backtick}{\backtick}{\backtick}}

\begin{document}

\title{Mutation-Guided LLM-based  Test Generation at Meta}


\author{Christopher Foster}
\affiliation{%
  \institution{Product Compliance and Privacy team, Meta Platforms}
  \city{Menlo Park}
  \country{USA}
}

\author{Abhishek Gulati}
\affiliation{%
  \institution{Product Compliance and Privacy team, Meta Platforms}
  \city{Menlo Park}
  \country{USA}
}

\author{Mark Harman}
\orcid{https://orcid.org/0000-0002-5864-4488}
\affiliation{%
  \institution{Product Compliance and Privacy team, Meta Platforms}
  \city{London}
  \country{UK}
}

\author{Inna Harper}
\orcid{https://orcid.org/0009-0008-9359-0949}
\affiliation{%
  \institution{Developer Infrastructure team, Meta Platforms}
  \city{London}
  \country{UK}
}

\author{Ke Mao}
\orcid{https://orcid.org/0000-0003-3956-9184}
\affiliation{%
  \institution{WhatsApp team, Meta Platforms}
  \city{London}
  \country{UK}
}

\author{Jillian Ritchey}
\affiliation{%
  \institution{Messenger team, Meta Platforms}
  \city{New York}
  \country{USA}
}

\author{Herv\'{e} Robert}
\affiliation{%
  \institution{Product Compliance and Privacy team, Meta Platforms}
  \city{Menlo Park}
  \country{USA}
}

\author{Shubho Sengupta}
\orcid{https://orcid.org/0009-0007-4204-5185}
\affiliation{%
  \institution{FAIR, Meta Platforms}
  \city{Menlo Park}
  \country{USA}
}

\renewcommand{\shortauthors}{Harman,  Sengupta}

\begin{abstract}
This paper\footnote{Author order is alphabetical. The corresponding author is Mark Harman.} describes Meta's \toolname~system for mutation-guided LLM-based test generation. 
\toolname~ generates relatively few mutants (aka simulated faults), 
compared to traditional mutation testing.
Instead, it focuses on generating currently 
undetected faults that are specific to an issue of concern. 
From these currently uncaught faults, \toolname~ generates tests that can catch them,
thereby `killing' the mutants and consequently  hardening the platform against regressions.
We use privacy concerns to illustrate our approach, but \toolname~can harden code against {\em any} type of regression.
In total, \toolname~ was applied to
10,795 Android Kotlin classes in \NumberOfProducts~ software platforms deployed by Meta, from which it generated 9,095  mutants and 571 privacy-hardening test cases.
\toolname~also deploys an LLM-based equivalent mutant detection agent 
that achieves a precision of 0.79 and a recall of 0.47 (rising to 0.95 and  0.96  with simple pre-processing).
\toolname~was used by   Messenger and WhatsApp test-a-thons where engineers accepted  \OverallAcceptRate~ of its tests, 
judging  \PercentRelevant~to privacy relevant.
We  conclude that \toolname ~hardens code against specific concerns and that, even when its tests do not directly tackle the specific concern, engineers find them useful for their other benefits.
\end{abstract}


\maketitle

\section{Introduction}
In this paper, we report on Meta's deployment of \toolname\footnote{\toolname~(\toolExpendedName)~is so-named because it is an automated test generation tool that `hardens' compliance with respect to chosen issues of concern.}, an agentic LLM-based tool for generating tests to target specific classes of faults. 
The paper focuses on automated privacy hardening: the problem of automatically generating unit tests to reduce the risk of future regressions with respect to privacy issues. 
However, the approach can be applied to any issue and is not confined solely to tackling privacy issues.
The paper reports results from the deployment of \toolname~ at Meta between \InitialDate~and \FinalDate.

Although there has been a great deal of recent attention on LLM-based test generation \cite{mhetal:TestGen-LLM,ryan2024code,chen2024chatunitest,schafer2023empirical,liu2024llm}, 
there has been little work on developing tests for specific classes of fault.
Many companies have exposure to specific high impact faults related to important issues such as security, integrity, and privacy.
The importance of such issues makes it equally important to have test generation techniques
that can target these specific classes of faults.

Organizations typically collect data about  bugs found during development.
This provides a  rich source of information with which to guide test generation.
The challenge, therefore, is to find a way to generate tests that target specific issues on the basis of this information. 
We believe mutation testing holds the key:
our key insight is to construct mutants that denote faults that are both relevant to the issue of concern and also currently not caught (unkilled) by any existing test case,
and to use these as prompts to LLM-based test generation. 
This results in an overall agenetic LLM-based workflow, 
in which agents essentially generate problem-specific `super bugs' and the tests that can catch them.

At Meta, we have developed and deployed just such an approach and tool, 
\toolname, for automated unit test generation.
{\toolname}'s  workflow is based on the principle of Assured LLM-based Software Engineering \cite{mhetal:intense24-keynote}. 
In Assured LLMSE, the goal is not merely to use language models to tackle software engineering challenges, 
but to automatically generate software engineering artifacts that come with {\em assurances}. 
In the case of  \toolname, the artifacts are tests and they come with the following assurances:

\begin{enumerate}
    \item {\bf Buildable}: The proposed new tests build, so are free from syntax errors and missing dependencies;
    \item {\bf Valid Regression Tests}:  The tests pass and do so consistently, so they are non-flaky~\cite{mhpoh:scam18-keynote} regression tests;
    \item {\bf Hardening}: The new tests catch faults that {\em no} existing test can catch;
    \item {\bf Relevant}: Many of the new tests are closely coupled to the issue of concern;
    \item {\bf Fashion Following}: Most of the new tests respect the coding style used by the existing tests available for the class under test; they are `fashion followers' \cite{mhetal:TestGen-LLM}.
\end{enumerate}

The first three assurances are boolean in nature; they are unequivocal {\em guarantees} made by \toolname~ about the tests it proposes to the engineer. 
The fourth and fifth are more aspirational and probabilistic in nature, reflecting the underlying (LLM) technology used to construct tests. 
The degree to which tests are relevant and `fashion follow' the existing tests is an inherently non-boolean and subjective attribute. 
As such, while the boolean criteria are provided as verifiable guarantees that \toolname~ promises to meet,   
the two more probabilistic assurances are best assessed through the standard code review process routinely undertaken by engineers.
Finally, when it finds that a newly-generated test adds coverage, 
\toolname~ also measures coverage achieved, including the result  in the assurances provided to the engineer reviewing the test.

The newly generated tests need not extend coverage, since they may find faults simply by covering existing lines of code in some new way; the same path with different data, for example.
It is well known \cite{mike:icse17} that mutation testing has this property, allowing mutation testing to claim superiority over  structural coverage test adequacy criteria such as line  coverage.
As shown in Section~\ref{sec:why_mutation}, our findings further underscore the importance of mutation testing in `going beyond' purely  structural test adequacy criteria. 

The paper focuses on privacy hardening, 
but we believe that our approach will find  many applications to software testing, more generally.
It combines existing well-known and widely-studied approaches to 
mutation testing, 
assured LLMSE, 
and test generation.
However, this new combination  allows it to tackle a wide range of other problems.
The potential reach and impact of this approach derives from the fact that it answers
a fundamental question for automated software test generation:

\begin{quote}
\it
    How can we automatically transform  vague, incomplete and even contradictory 
    textual narratives about software concerns, into unit tests that guard against bugs that, 
    if introduced, 
    may yield field failures that manifest these concerns?
\end{quote}

This fundamental nature makes the approach  very widely applicable: we can use it in any situation where we have a way to capture, even in the most vague of textual descriptions, concerns we  have about potential issues our systems will face.
Furthermore, because \toolname~ generates simulated faults as an intermediate stage in generating tests, we can use the distributions of simulated fault prevalence  over the code base as a proxy for the risk exposure of individual system components to the issue under test.
Finally, through simulation \cite{jaetal:ease21-keynote}, we can further assess the likely impact (or severity) of such faults, were they to  manifest as field failures.

There are four primary contributions of the paper:


\noindent{\bf Empirical results}:
  We report results   from the application of \toolname~ to \NumberOfProducts~ real world software platforms deployed by Meta.
    In total, \toolname~generated 4,660 candidate mutants (simulated privacy faults) that it `believed' to be potentially killable 
    from 10,795 classes under test.
    From these candidates, \toolname~ was able to generate an additional 571 unit tests.
    Of these 571 tests, 277 would have been discarded had we chosen to focus solely on the line coverage test adequacy criterion, 
    underlining the importance of mutation testing over such  coverage.

\noindent{\bf Deployment Experience}:    
We report  both qualitative and quantitative outcomes from the development, deployment and evaluation of  \toolname~ at Meta in 2024. 
Overall, the tests automatically generated by \toolname~achieved an  acceptance rate of \OverallAcceptRate~ from the engineers who reviewed them, with \PercentRelevant~ being judged privacy relevant.

\noindent{\bf Equivalent Mutant Detection}:    
We present an  evaluation of the {\toolname} equivalent mutant detection agent's ability to detect equivalent mutants, which yields precision and recall of 0.79 and 0.47 respectively. 
    The figure for precision  and recall rises to 0.95 and  0.96 respectively, 
    when the  agent is combined with simple pre-processing. 
    
\noindent{\bf Lessons from Industrial Application}:    
We detail the lessons learned,  and open problems and research challenges raised by this application of Assured LLMSE to mutation-based  test generation.

\section{The \toolname~System}
\begin{figure*}

\centerline{\includegraphics[width=1.25\linewidth, trim={1cm 0 0 5cm}, clip]{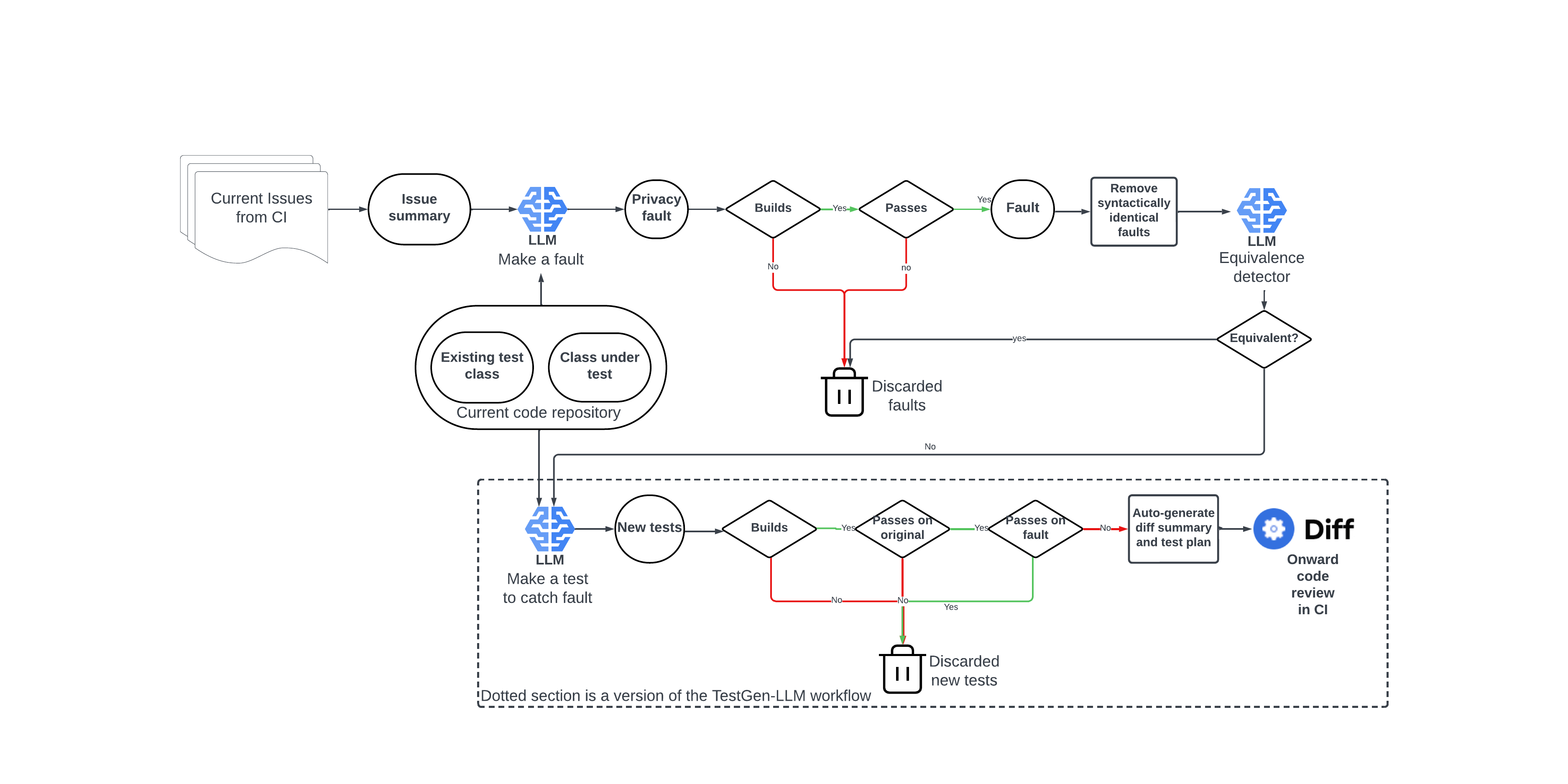}}
\vspace{-14mm}
\caption{Top level architecture of the principal \toolname~components. 
The dotted section denotes a (slightly modified) version of the TestGen-LLM tool, on which we previously reported \cite{mhetal:TestGen-LLM}.
The workflow preceding this, shown above in the figure, is the additional agenetic workflow for generating candidate faults to drive the generation of tests. Solid rectangles denote components that are fully automated but entirely rule-based (and therefore do not use LLMs).}

\label{fig:top_level}
\end{figure*}

Figure~\ref{fig:top_level} presents the overall architectural pipeline of the \toolname~System.
\toolname~starts with free from text about an issue of concern.
This textual input could come from one or more of a variety of sources, including (but not limited to):

\begin{enumerate}
    \item Previous faults found in development;
    \item User requirements;
    \item System technical constraints;
    \item Concerns raised by engineers, managers or organizational leadership about undesirable system behaviours;
    \item Regulatory requirements set by legislative bodies and compliance enforcement organizations.
\end{enumerate}

This is the sense in which \toolname~ is a `compliance hardener': it improves (`hardens') the ability of the deployed regression test infrastructure to detect regressions that might lead to non-compliance with respect to the issue of concern.

The results presented in this paper were obtained by using  privacy hardening concerns from previous faults found in development.
The single language model Llama 3.1 70Bn~\cite{llama31} was used in all the agents reported on.
The prompts used by the three LLM agents from Figure~\ref{fig:top_level} can be found in Table~\ref{tab:prompts}.

\begin{table*}
\caption{The  three simple prompts used in the architecture diagram in Figure~\ref{fig:top_level}}
\vspace{-4mm}
\label{tab:prompts}
\scriptsize
\begin{tabularx}{\textwidth}{|p{2.3cm}|X|}
 \hline
Agent  name & Prompt Template, in which \{ $\cdots$ \} denotes a parameter to the template\\
\hline 
Make a fault        &  {\tt CONTEXT: \{context\_about\_concern\} INSTRUCTION:
                       Here is a Kotlin class and a test class with some unit tests for the class under test {\triplebacktick}\{class\_under\_test\}{\triplebacktick}. {\triplebacktick}\{existing\_test\_class\}{\triplebacktick}. 
                       Write a new version of the class under test in which each method is replaced by a new version of that method that contains a typical bug that introduces a privacy violation similar to \{diff\}.  
                       Delimit the mutated part using the comment-pair {\backtick}// MUTANT <START>{\backtick} and {\backtick}// MUTANT <END>{\backtick} }\\
Equivalence detector           &  {\tt I'm going to show you two slightly different versions of a  Kotlin class. Here is the first version of the Kotlin class:{\triplebacktick}{class\_version1}{\triplebacktick}. Here is the second version of the Kotlin class:{\triplebacktick}{class\_version2}{\triplebacktick}. 
INSTRUCTION: If the first version of the class will always do exactly the same thing as the second version of the class, 
just respond with {\backtick}\{yes\}{\backtick}. However, if the two versions of the class are not equivalent, respond with {\backtick}\{no\}{\backtick}, 
and give an explanation of how execution of the first version can produce 
a different behaviour to execution of the second version.}\\
Make a test to catch fault           &  {\tt What follows is two versions of a Kotlin class under test. An original correct class and a mutated version of that class that contains one mutant per method, each of which represents a bug. Each bug is delimited by the comment-pair {\backtick}// MUTANT <START>{\backtick} and {\backtick}// MUTANT <END>{\backtick}. The original class and its mutant are followed by a test class that contains unit tests for the original correct class under test. This is the original version of the class under test:{\triplebacktick}\{original\_class\}{\triplebacktick}. This is the mutated version of the class under test:{\triplebacktick}\{mutated\_class\}{\triplebacktick}. Here is the existing test class:{\triplebacktick}\{existing\_test\_class\}{\triplebacktick}. Write an extended version of the test class that contains extra test cases that will fail on the mutant version of the class, but would pass on the correct version.}\\
\hline
\end{tabularx}
\end{table*}

Our focus is the development, deployment and evaluation of  mutation-guided agenetic workflows.
We have not yet felt the need to extend to more sophisticated prompting, nor to use fine-tuning, nor to exploit language model ensembles \cite{mhetal:TestGen-LLM}, all of which would improve on the results we report. 
Therefore, our results denote only a baseline against which to measure futures developments.
Nevertheless, despite these limitations,  we believe the results clearly highlight the potential for significant advances in test generation; its applicability and effectiveness. 
We would be interested and excited to collaborate with the wider research community, and hope this paper stimulates further work in this area.

In the \toolname~workflow depicted in Figure~\ref{fig:top_level}, 
the issue summary generates prompts that stimulate another LLM-based agent to
generate faults; 
walking the code base, using existing tests, classes under test, and the summary as prompt ingredients.
The prompts instruct the LLM to generate simulated faults (aka mutants \cite{yjmh:analysis}).
There is no guarantee that these mutants will change the code under test, because 
it may not be relevant to the issue being hardened. 

However, even when the fault generator does change the syntax, it may not change the semantics; the mutants could be equivalent  mutants \cite{xymhyj:equivalent}.
To tackle this equivalent mutant problem, \toolname~uses a further agent, the Equivalence Detector agent.
This Equivalence Detector agent uses the  LLM-as-judge approach~\cite{gu2024survey,zheng2023judging}.
The judge is instructed to determine whether the mutant is equivalent to the original class under test.
Although the underlying problem of mutant equivalence is undecidable \cite{yjmh:analysis}, 
we found that this approach was surprisingly effective due to the properties of the mutants so-generated (Section~\ref{sec:equivalance} contains details).

The faults generated by the initial phase of the workflow are used to generate prompts for test generation.
The test generation phase shown in a dotted box is a slightly modified version of the previously published
TestGen-LLM tool \cite{mhetal:TestGen-LLM}.

\section{Results from  Meta's \toolname~Deployment}
We applied \toolname~to the \NumberOfProducts~   Meta platforms \ProductList.
Facebook Feed and Instagram are social media platforms. 
WhatsApp and Messenger are messaging platforms. 
Oculus is a set of virtual reality headsets.
Wearables are platforms for augmented reality glasses and wrist controllers.
We also included `cross-app' software products that provide features affecting more than one of these platforms.

Table~\ref{tab:mutants_made} presents top level summary statistics for the application of \toolname~ to these platforms.
Once a mutant is found that builds and passes, the workflow terminates\footnote{
This decision was motivated by the use case: we first want to establish whether a test is relevant.
When we find a test that is relevant, we can choose to focus in on  the code it tests, hoping that adjacent code will also have a higher likelihood of relevance.} for the current class under test.
The number of candidate mutants reported in Table~\ref{tab:mutants_made} is therefore bounded above by the number of classes under test.


By contrast with traditional rule-based mutation testing tools~\cite{coles:pit,yjmh:milu,schuler:javalanche,just:major},  \toolname~ generates relatively few, highly specific mutants, by design. 
Furthermore, the generated mutants have  a higher probability of relevance to the issue of concern than can be achieved by 
rule-based approaches.

As can be seen from Table~\ref{tab:mutants_made}, \toolname~ generates 9,095 mutants that build and pass 
from 10,795 classes under test.
Although the language model approach generates fewer and more specific mutants than rule-based approaches, 
we found that it also generates more equivalent mutants. 
For example, as Table~\ref{tab:mutants_made} reveals, 25\% of the mutants generated are trivially syntactically equivalent.
This contrasts with overall equivalent mutant generation rates that are typically around 10\% to 15\% for rule-based approaches \cite{papadakis:trivial,xymhyj:equivalent}.
Fortunately, as revealed in  Section~\ref{sec:equivalance}, the equivalent mutant problem is relatively unimportant for our use case.

\begin{table*}
\caption{
Meta platforms used to evaluate \toolname~and the numbers of  \PrivacyMutants~ `believed' non-equivalent by the equivalence detector agent from Figure~\ref{fig:top_level}. 
Percentages in the final four columns are distributions of {\toolname}'s equivalence `belief' over mutants that build and pass, while 
those in the fourth column report the percentage of all mutants that build and pass. 
}
\vspace{-4mm}
  \centering
  \begin{tabular}{|l|r|r|r||r|r|r|r|}
    \hline
    Platform & Number of    & Number of       & Number of      & \multicolumn{4}{c|}{Number of mutants that build and pass  and ...}             \\
            & classes      & mutants         & mutants        & are             & the agent       & for which      & the agent     \\
            & under test   & generated       & generated      & syntactically   & believes        & the agent      & believes      \\ 
            &              & in total        & that build     & identical       & to be           & gives no       & to be non-    \\
            &              &                 & and pass       &                 & equivalent      & answer          & equivalent    \\
    \hline
    Facebook Feed &   346  &       1,097     &    252 (23\%)  &       50 (20\%) &      19 (7.5\%) &     19 (7.5\%) &   164 (65\%)  \\
    Messenger     & 3,339  &       9,185     &  2,922 (32\%)  &      742 (25\%) &     384 (13\%)  &    401 (14\%)  & 1,395 (48\%)  \\
    Instagram     & 1,691  &       5,199     &  1,381 (27\%)  &      277 (20\%) &     154 (11\%)  &    208 (15\%)  &   742 (54\%)  \\
    Aloha         &   175  &         576     &    144 (25\%)  &       32 (22\%) &      16 (11\%)  &      6 (4\%)   &    90 (63\%)  \\
    Wearables     & 2,841  &       8,592     &  2,468 (29\%)  &      621 (25\%) &     207 (8\%)   &    326 (13\%)  & 1,314 (53\%)  \\
    Cross-app     &   805  &       1,819     &    562 (31\%)  &      146 (26\%) &      64 (11\%)  &     82 (15\%)  &   270 (48\%)  \\
    Oculus        &   325  &         997     &    279 (28\%)  &       96 (34\%) &      17 (6\%)   &     16 (6\%)   &   150 (54\%)  \\
    WhatsApp      & 1,273  &       4,212     &  1,087 (26\%)  &      282 (26\%) &     155 (14\%)  &    115 (11\%)   &  535 (49\%)  \\
    \hline
    \hline
    Totals       & 10,795  &      31,677     &  9,095 (29\%)  &    2,246 (25\%) &   1,016 (11\%)  &   1,173 (13\%)  & 4,660 (51\%) \\
    \hline
    \hline
  \end{tabular}

  \label{tab:mutants_made}
\end{table*}

Of the 9,095 mutants that build and pass, 4,660 (51\%) were deemed to be non-equivalent by the workflow, 
and thus become the subject of each of the prompts used for test generation.
The test generation phase uses similar prompts to those reported previously~\cite{mhetal:TestGen-LLM}, but based on covering the mutant rather than covering uncovered lines.

\section{Engineers' Evaluation of \toolname}
We followed the tried-and-tested formula \cite{mhetal:TestGen-LLM} for the deployment of a new software testing technology at Meta, starting with initial trials and moving onto to  test-a-thons led by engineers, thereby evaluating initial deployment.

\subsection{Initial Trial}
At Meta, a code change submitted to the Continuous Integration (CI) system  is called a `diff' (short for differential).
In order to get initial feedback from engineers on the tests generated by the first version of \toolname, 
we used it to generate 30 new test cases, submitting each as a separate diff for review.
The generated tests were recommended to engineers in the same way as any other code modification, such as those created by  human engineers. 
That is, \toolname~ submits tests, as diffs,  through the normal review process.
The \toolname~diff summary explains the kind of fault caught by the test case giving, as  a  specific example, the  mutant that \toolname~ has already determined to be killed by the test case.


We wanted to obtain broad coverage of Meta's platforms, so included 
messaging apps like WhatsApp and Messenger, 
traditional social media platforms such as Facebook Feed, 
as well as hardware technologies 
such as Oculus and Wearable Computing, 
and also code that spanned over multiple apps.

Table~\ref{tab:initial} presents details of these 30 diffs.
As can be seen, the overall acceptance rate is 
90\% accepted of those submitted (27 of 30), 
and 93\% of all diffs reviewed (27 of 29).
Initial results for acceptance rate were deemed  to be highly encouraging, so we proceeded to 
build out the Minimal Viable Product (MVP) and use it in the test-a-thons reported in Section~\ref{sec:test-a-thons}.

The primary feedback from developers who reviewed these 30 diffs was as follows. Engineers reported that

\begin{enumerate}
    \item  Many of the tests added coverage, which they computed themselves manually. 
    They asked that coverage information be automatically computed and reported in the diff summaries, as well as the faults found.
    \item Some of the tests were relevant to privacy, and could be useful in hardening against future privacy regressions. 
    They also felt that having the specific example faults was very helpful in understanding the behavior of the tests. 
    \item Even for those which were not clearly related to privacy, the tests seemed to be adding  value by tackling corner cases or adding coverage.
\end{enumerate}




\begin{table}
  \caption{Results from initial trial \toolname~ deployment on 6 platforms and cross-app products to gauge developers' reactions, gain feedback and guide development.} 
  \vspace{-4mm}
\small
  \centering
  \begin{tabular}{|l|r|r|r|r|r|}                                                       
    \hline
    Platform &   Number             & \multicolumn{4}{c|}{Status after human review is ...}   \\
            &   of tests            & Accepted    & Accepted & Rejected & Not                      \\
            &               & `as is' by  & with     & by the   &  reviewed                           \\
            &               & the         & simple   & engineer &                        \\
            &               & engineer    & changes  &          &                        \\
    \hline
    FB Feed     &  4   & 4    & 0    &  0   & 0 \\
        \hline
    Messenger         & 12   & 8    & 3    &  1   & 0 \\ 
        \hline
    Aloha         &  2   &  2   & 0    &  0   & 0 \\
        \hline
          Wearables     &  3   &  1   &  0   &  1   & 1 \\
        \hline
         Cross-app     &  2   &  2   &  0   &  0   & 0 \\
        \hline
        Oculus        &  3   &  3   &  0   &  0   & 0 \\
       \hline
        WhatsApp      &  4   &  3   & 1    &  0   & 0  \\
     \hline
        \hline
        Totals        & 30   &   23 & 4  &2 &  1 \\
                \hline
  \end{tabular}

  \label{tab:initial}
\end{table}

\subsection{Experience from Privacy Test-a-thons}
\label{sec:test-a-thons}
In the week of 9th December 2024, we conducted two test-a-thons,  
focusing on the application of \toolname~to Meta's two messaging platforms: WhatsApp and Messenger.

As with the initial trial, test cases were submitted as diffs into the normal continuous integration review process.
However, the diff summary additionally claimed additional coverage, for those tests that did add coverage as well as finding currently uncatchable faults.
The engineers participating in the test-a-thons reviewed the diffs for usefulness in the normal way they would any other diff, ultimately determining whether the diff is accepted, and thus lands into production.

\begin{figure}
\centerline{\fbox{\includegraphics[width=0.45\textwidth]{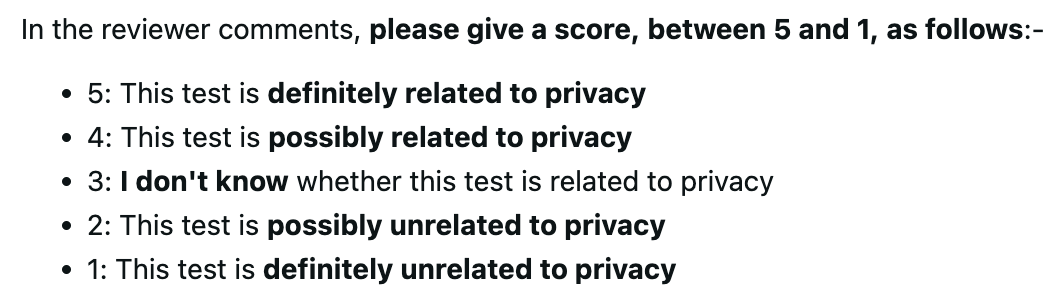}}}
\vspace{-1em}
\caption{Likert scale instructions given to code reviewers to score test cases according to their privacy relevance. }
\label{fig:likert}
\end{figure}

In order to evaluate the privacy relevance of each test, 
we additionally asked the software engineers to give each a score on a Likert scale~\cite{jebb:likert-review}.
Figure~\ref{fig:likert} is a screen capture of the exact instructions given to the engineers regarding this privacy relevance scoring procedure.

\noindent
{\bf Procedure for Messenger two-phase test-a-thon}:
The Messenger test-a-thon was conducted in two phases. 
The same 6 reviewers were used for both the first and the second phase.
All Messenger reviewers had strong expertise in testing, and (at least a) basic experience with privacy engineering.

In the first phase, 50 generated tests were selected from a randomly selected pool of 60 tests.
All 60 were prescreened manually and, as a result, 10 were excluded 
because they concerned only faults relating to Null Pointer Exceptions, two of which improved coverage, and eight of which did not.
These were excluded because we initially (wrongly, as it turned out) thought 
that such exception-catching tests would be rejected by engineers due to being irrelevant to privacy.
However, the pre-screening approach was abandoned based on experience from this first phase: 
Engineers proved ready to accept tests, 
even when they were not directly relevant, because they perceived other benefits.
Therefore, in the second phase, a further 50 tests were selected at random, without pre-screening.

\noindent
{\bf Procedure for the WhatsApp test-a-thon}:
In the WhatsApp test-a-thon, 72 tests\footnote{The decision to use 72 tests was based on the initial request from the 6 engineers that they could reasonably review 12 tests; this number being deemed large enough for calibration, yet  small enough to avoid reviewer overload.} were selected randomly from a pool of 120 available and allocated to 6 engineers to review.
No pre-screening was performed.
All 6 engineers had a background in both  privacy engineering and software testing and, as such, were well-placed with relevant expertise and highly calibrated in their expectations about privacy concerns.

Having completed the reviews of the initial 72 tests allocated, 
several of the engineers requested additional tests to review, having found the experience sufficiently rewarding.
Additional tests were therefore allocated at random from the remaining pool of 48 available.
Therefore, in total, 91 WhatsApp tests were reviewed for usefulness, and 90 were reviewed for relevance.


\noindent
{\bf Results from the test-a-thons}:
Table~\ref{tab:testathon} presents summary statistics for the two test-a-thons.
The upper table gives the  number of tests reviewed for usefulness and for privacy relevance
and, of these, the number that were accepted and rejected for usefulness, and scored for relevance.
The lower table gives the proportions of tests reviewed for usefulness that were accepted and rejected, 
and the proportion of those reviewed for relevance that fall into each of the five categories of the Likert scale on which relevance was assessed by the software engineers.
Percentages are rounded to the nearest whole number percentage point, and so may not quite total 100\% due to rounding.

\begin{table*}
  \caption{Results from WhatsApp and Messenger Test-a-thons, where \toolname~was deployed and evaluated in December 2024.}
  \vspace{-4mm}
  \centering
  \begin{tabular}{||l|r|r|r||c||l|r|r|r|r|r||}
\hline
\hline
                                     & total number      & \multicolumn{2}{c||}{total number}   &~~&  total number        & \multicolumn{5}{c||}{}                    \\
                                     & of tests          & \multicolumn{2}{c||}{of tests }      &~~&  of tests            & \multicolumn{5}{c||}{privacy relevance}   \\
                                     & reviewed          & \multicolumn{2}{c||}{that were ...}  &~~&  reviewed            & \multicolumn{5}{c||}{scored at level ...} \\
{\bf Test-a-thon}                    & for {\bf usefulness}    & accepted      & rejected      &~~&  for {\bf relevance} & ~5~  & ~4~  & ~3~  & ~2~  & ~1~          \\
\hline
WhatsApp                             &  91               &  50                 &  41           &~~&  90                  &  6   & 29   &  8   & 24   & 23           \\
Messenger Phase 1 (pre-screened)     &  50               &  47                 &   3           &~~&  44                  &  5   &  8   &  5   &  4   & 22           \\
Messenger Phase 2 (not pre-screened) &  50               &  43                 &   7           &~~&  41                  &  4   & 11   &  0   &  7   & 19           \\
\hline
Overall Total                        & 191               & 140                 &  51           &~~& 175                  & 15   & 48   & 13   & 35   & 64           \\

\hline
\hline
\multicolumn{11}{l}{}\\ 
\multicolumn{11}{l}{Proportions of tests in each category for usefulness and relevance :-}\\  
\hline
\hline
Over all test-a-thons                &                   & \OverallAcceptRate  &  27\%         &~~&                      &  9\% & 27\% &  7\% & 20\% & 37\%         \\
\hline
\hline
WhatsApp                             &                   & 56\%                &  44\%         &~~&                      &  7\% & 32\% &  9\% & 26\% & 26\%         \\
Messenger Phases 1 and 2             &                   & 90\%                &  10\%         &~~&                      & 11\% & 22\% &  6\% & 13\% & 48\%         \\
\hline
Messenger Phase 1 alone              &                   & 94\%                &   6\%         &~~&                      & 11\% & 18\% & 11\% &  9\% & 50\%         \\
Messenger Phase 2 alone              &                   & 86\%                &  14\%         &~~&                      & 10\% & 27\% &  0\% & 17\% & 46\%         \\
\hline
\hline
  \end{tabular}

  \label{tab:testathon}
\end{table*}

As Table~\ref{tab:testathon}  reveals, the overall acceptance rate was \OverallAcceptRate.
This is very similar to acceptance rates from previous test-a-thons that focused exclusively on elevating coverage \cite{mhetal:TestGen-LLM}.
Overall, about one third of the tests deemed definitely {\em irrelevant} to privacy.
This makes it all the more interesting that the rejection rate is much lower than this, at 27\%.
Indeed, only approximately \PercentRelevant~were  deemed to be either possibly or definitely related to privacy.
We therefore observe that the engineers are very willing to accept tests which may not be relevant to their current concern, 
when they are found  useful for other reasons.

In looking at the individual comments left by the engineers on each of the 191 tests that were reviewed for usefulness, we observe, more anecdotally, 
 that the engineers were likely to accept tests for two primary reasons:

\begin{enumerate}
    \item They add line or branch coverage of non-trivial code
    \item They covered a tricky corner case, such as handling special values, even when this failed to add coverage
\end{enumerate}

Tests that were rejected tended  not to add coverage, or to add coverage only of trivial code, such as one-line behavioral functions.
Tests were also rejected if they 
were written in a style that was deemed to be  unsuitable, such as using a deprecated API, something that was also observed in previous work~\cite{mhetal:TestGen-LLM}.

In terms of the relevance of the tests, we believe that the finding that \PercentRelevant~ are deemed relevant to privacy is a positive overall outcome.
This is because we devoted so little of the overall process to weeding out tests that could be automatically determined  to be irrelevant.
As such, the result for relevance can be considered a baseline for comparison with future work.
We believe that with additional static analysis, and further LLM-as-judge agents in the overall workflow, it could be considerably improved.

We also noticed that the comments about relevance for tests scored 4 and 2 were often quite similar, indicating uncertainty, and the belief that the test may be relevant to privacy, but the engineer was not certain.
Therefore, an upper bound on those potentially relevant to privacy is approximately two thirds of those tests assessed.
Given the  \OverallAcceptRate~ acceptance rate, we conclude that \toolname~ adds privacy hardening in about one third  of the cases, and does not cause unnecessary developer friction in considering the remaining cases.


There is an interesting difference in the relevance score between the two phases of the Messenger test-a-thon results reported in Table~\ref{tab:testathon}. 
In the first phase, the engineers stated that they were unsure in 10\% of the cases, but this percentage dropped to zero in the second phase.
This apparent growing scoring `confidence' indicates a potential `learning effect', 
which has been seen in similar empirical studies of software engineers during multiphase trials \cite{jdmhmolh:tse}.

Finally, looking at the differences between the acceptance rates and relevance assessments for WhatsApp and Messenger, we also see interesting differences.
WhatsApp engineers rated the tests to be at least as relevant to privacy (39\% for WhatsApp vs. 33\% for Messenger), yet accepted fewer tests (56\% WhatsApp vs. 89\% for messenger).
All tests that were acceptable for usefulness were landed into production.
We believe that the different acceptance rates may simply denote different cultures between different teams; deciding to land a test into production is an inherently subjective judgment and may be influenced by team culture.

\section{The Equivalent Mutant Problem}
\label{sec:equivalance}
All approaches to  mutation testing need to tackle the problem of equivalent mutants \cite{yjmh:analysis}:
the mutant may be {\em syntactically} 
different to the original program, but we cannot guarantee it will be {\em semantically} different,
because the underlying program equivalence problem is  undecidable \cite{xymhyj:equivalent,van2021mutantbench}.

There are a number of techniques in the literature that can  weed out {\em some} of the equivalent mutants~\cite{madeyski2013overcoming}.
However, the undecidability of the problem means that some equivalent mutants will inevitably remain, so
engineers and automated test generators may waste  time trying to kill (unkillable) equivalent mutants.

\subsection{Equivalent Mutants Have no Impact on   \toolname}
For the application of mutation-guided test generation, 
the  equivalent mutant problem has no direct impact on engineers.
Our workflow requires  only that engineers review test cases, not mutants.
The only scenario in which an engineer {\em might} consider  looking at a mutant, 
would be  to see an example of the kind of faults that can be caught by the test case they are reviewing. 
By construction, such mutants are {\em non-equivalent}, so the engineer will {\em never} see an equivalent mutant.
This relegates the equivalent mutant problem to a relatively subordinate position in our overall use case for mutation testing. 

Nevertheless, it would be  inefficient to generate many equivalent mutants, 
because \toolname~ would waste computational resources trying to kill the unkillable. 
We therefore incorporate, into the agentic workflow, 
an LLM-based agent that checks for mutant equivalence.
The remainder of this section reports on the evaluation of the effectiveness of this agent.

\subsection{Detecting Equivalent Mutants}
To evaluate the performance of the equivalence  detector agent from Figure~\ref{fig:top_level}, 
we performed a manual study on a random selection of mutants drawn from the four platforms with the most mutants available.
The purpose of this manual analysis is to answer the research question:

\begin{quote}
     How good is the Equivalence Detector Agent?
\end{quote}

We must manually analyze mutants to answer this research question, because it requires a ground truth for equivalence. 
However, it is important to underline that it is {\em never} necessary for a {\em practicing software engineer} to manually evaluate {\em any} mutant.


For each of the four platforms, we attempted to sample 100  mutants that would still build and pass on the original code.
The actual number manually analyzed for each   sample was slightly different to the intended 100, 
because some code may have changed since the mutants were constructed.
This is the Mutant Relevance Problem~\cite{ojdanic:keeping}: previously generated mutants may no longer be relevant when the code  changes after they were constructed.

Mutant relevance does not impact the deployment of \toolname, 
since \toolname~ generates mutants on the fly, discarding them once tests have been generated from them.
However, it did have a minor impact on our mutant sampling to evaluate the research question:
Since we cannot be sure mutants in changed code remain valid at manual inspection time, 
we discarded them from the human analysis sample.

In our evaluation of equivalence detection, 
we are concerned only with so-called `weak' mutation testing, 
not `strong' mutation testing \cite{yjmh:analysis}. 
That is,  because \toolname~is a unit test generation technology, 
we do not need to consider  failed error propagation \cite{kaetal:analysis}, 
which would be required for strong mutation testing.
Failed error propagation is the situation in which 
mutant execution changes the local computation state, 
yet this change is always masked along 
every path to an observable output. 
In weak mutation testing, the mutant is non-equivalent if the local state is changed, 
irrespective of the whether the change propagates.
Manually detecting failed error propagation is a highly labor-intensive process, 
so we are fortunate that it is not required.

Table~\ref{tab:all_different_types_of_mutants_in_it_look_i_want_that_one_for_my_Christmas} describes the types of mutant, based on our manual analysis.
In this analysis, the terms `deletion' and `injection' each refer to semantic changes where the sole change is, respectively, the addition of values or the removal of values. 
The term `Injection and deletion' refers to cases where the semantics are changed by an independent combination of addition and removal. 
The category `other' covers cases where the change is more nuanced.
For example replacing one variable name with another, which could be counted as a deletion and an injection, but where the two are not independent.

In Table~\ref{tab:all_different_types_of_mutants_in_it_look_i_want_that_one_for_my_Christmas},
the column `equivalence is not obvious' counts the number of cases where the human assessor had to think noticeably longer 
about whether the mutant was equivalent. 
For all other cases, the decision was obvious.
As  Table~\ref{tab:all_different_types_of_mutants_in_it_look_i_want_that_one_for_my_Christmas} reveals, the decision is `obvious' for a surprisingly large number of mutants.

\begin{table*}
\caption{Number of mutants of different semantic types generated. 
We used human analysis to determine ground truth equivalence, and the different types of semantic transformation used to generate a mutants.}
\vspace{-4mm}
\centering
\begin{tabular}{|l|r|r|r|r|r|r|r|r|r|}
\hline
\hline
                   &          & \multicolumn{3}{c}{mutant equivalence}      &   \multicolumn{5}{|c|}{semantic type of mutation transformation}  \\     
Platform            & Total    &  Ground      & LLM agent    & Equivalence  & Deletion     & Injection  & Deletion   & Misleading  & Other       \\      
                   & mutants  &  truth        & claims       & is not       &              &            & and        & comment     &             \\
                   & analyzed &  equivalent   & equivalence  & obvious      &              &            & injection  &             &             \\
\hline
Messenger          &  91      &  38 (42\%)    & 27 (30\%)    &    2 (2\%)   &  23   (25\%) & 32 (35\%)  & 19 (21\%)  & 11 (12\%)   & 6  (7\%)    \\
Wearables          & 101      &  32 (32\%)    & 11 (12\%)    &    3 (3\%)   &  26   (26\%) & 24 (24\%)  & 20 (20\%)  & 24 (24\%)   & 7  (7\%)    \\ 
WhatsApp           &  99      &  30 (30\%)    & 16 (16\%)    &    3 (3\%)   &  29   (29\%) & 26 (26\%)  & 16 (16\%)  & 26 (26\%)   & 2  (2\%)    \\
Instagram          &  90      &  37 (41\%)    & 12 (13\%)    &    2 (2\%)   &  27   (30\%) & 14 (16\%)  & 11 (12\%)  & 32 (36\%)   & 6  (7\%)    \\
\hline
\hline
Overall Totals     & 381      & 137 (36\%)    & 66 (17\%)    &   10 (2.6\%) & 105   (28\%) & 96 (25\%)  & 66 (17\%)  & 93 (24\%)   & 21 (6\%)    \\
\hline
\hline
\end{tabular}

\label{tab:all_different_types_of_mutants_in_it_look_i_want_that_one_for_my_Christmas}
\end{table*}

The category `Misleading comment' is perhaps the most interesting category. 
These are cases where the mutant is clearly equivalent, 
because all the mutant generation agent has done is to insert comments; no executable code has been added.
An example of such a misleading injected comment is:  
``{\tt
// Introduce a bug by not checking if the user is inactivated before resetting their custom reactions}''.
Since  language models are text-based predictive technologies that currently do not specifically distinguish executable code from comments, such misleading comments are clearly problematic.
They can easily mislead the equivalence detector agent 
into predicting that a mutant is {\em not} equivalent, when it clearly {\em is} equivalent.

The reason that this category is so interesting, 
is that  \toolname~could simply strip out any comments added when constructing a mutant before submitting it to the equivalence detector agent.
This simple pre-processing step would render all cases of misleading comments syntactically identical.
They would then be automatically discarded without even needing to consult the Equivalence Detector Agent.
Based on the evaluation reported here, we decided to incorporate this pre-processor into {\toolname}'s workflow.

\subsection{Equivalence Detector Precision and Recall}
Table~\ref{tab:equivalence} reports the overall precision and recall of the equivalence detector agent.
The detailed results for each platform are shown in Table~\ref{tab:equivalence_each}.
As Table~\ref{tab:equivalence} reveals, when we consider `unsure' as the same as `equivalent', 
the precision is good, at 0.79 over all platforms studied. 
However, recall is relatively low, at 0.47.
If we wanted even greater precision we could treat `unsure' as `non-equivalent', since this gives precision of 0.97 (and a recall of 0.44).

Based on this precision and recall, we can have high confidence that, when the detector determines a mutant 
to be equivalent, it is very likely to be correct. 
However, it weeds out only approximately half of the equivalent mutants.
High precision means \toolname~will not discard many non-equivalent mutants. 
Not losing mutants means not losing the tests that \toolname~ might generate from them.
By contrast, higher recall would merely save some computational resource. 

As such, 
high precision is generally more valuable than high recall for our use case, 
because we are prepared to spend computational resources to automatically generate good unit tests.
However, the  lower the recall, the more often  \toolname~will inefficiently seek to `kill the unkillable',
so we still want the highest recall achievable without reducing precision.

As Table~\ref{tab:mutants_made} shows, 
approximately 25\% of all mutants  are trivially equivalent, 
being syntactically identical to the original.
Furthermore, Table~\ref{tab:all_different_types_of_mutants_in_it_look_i_want_that_one_for_my_Christmas}
reveals that
61\% of all equivalent mutants are almost trivially equivalent, 
because they contain only a single `misleading' comment.
Syntactically identical mutants are removed by a simple lexical comparison, while
a simple pre-processing transformation to remove comments would remove the `misleading comments' category.
Therefore, we can combine the LLM-based detector with a simple rule based pre-processor to yield an 
overall precision and recall of 0.95 and 0.96 respectively
(See Table~\ref{tab:equivalence}).

These figures for precision and recall are surprisingly high, 
given that the underlying problem is undecidable.
However, we cannot claim that these surprisingly good results arise
because the mutant equivalence detector agent is excellent at determining program equivalence.
Rather, they are more reflection of the kind of mutants the detector needs to  judge for equivalence.
That is, the mutant generation agent tends to make changes that either  obviously introduce semantic changes, 
or are obviously equivalent (such as those that only change comments).


\begin{table}
\caption{Precision \& recall for the equivalence detector agent}
\vspace{-4mm}
\centering
\begin{tabular}{|r|r|r|r||r|r|}
\hline
\multicolumn{6}{|c|}{{\bf when unsure is simply not counted}}\\
TP & FP & TN & FN & Precision & Recall\\
56 & 2  & 205 & 72 & 0.97 & 0.44 \\
\hline

\hline
\hline
\multicolumn{6}{|c|}{{\bf with unsure counted as  equivalent}}\\
TP & FP & TN & FN & Precision & Recall\\
65 & 17 & 205 & 72 & 0.79 & 0.47\\
\hline

\hline
\hline
\multicolumn{6}{|c|}{{\bf with identical code counted as equivalent}}\\
TP & FP & TN & FN & Precision & Recall\\
161 & 17  & 205 & 72 & 0.90 & 0.69 \\
\hline

\hline
\hline
\multicolumn{6}{|c|}{{\bf with the stripping of all added comments}}\\
TP & FP & TN & FN & Precision & Recall\\
183 & 9 & 251 & 8  & 0.95 & 0.96\\
\hline
\end{tabular}

\label{tab:equivalence}
\end{table}

\begin{table*}
\caption{Precision (Prec.) and Recall achieved by the equivalence detector agent for the four  platforms with most mutants}
\vspace{-4mm}
\centering
\scriptsize
\begin{tabular}{||r|r|r|r||r|r|||r|r|r|r||r|r|||r|r|r|r||r|r|||r|r|r|r||r|r||}
\hline
\multicolumn{6}{||c|||}{{\bf Messenger Mutants}} & \multicolumn{6}{c|||}{{\bf Wearables Mutants}} & \multicolumn{6}{c|||}{{\bf WhatsApp Mutants}} &\multicolumn{6}{c||}{{\bf Instagram Mutants}} \\
\hline
\hline
\multicolumn{24}{||c||}{}\\
\multicolumn{24}{||c||}{{\bf When the agent is `unsure', this is simply not counted in determining precision and recall}}\\
TP & FP & TN & FN & Prec. & Recall   & TP & FP & TN & FN & Prec. & Recall   & TP & FP & TN & FN & Prec. & Recall   & TP & FP & TN & FN & Prec. & Recall   \\
17 & 2 & 50 & 19 & 0.89 & 0.47           & 11 & 0 & 55 & 18 & 1.0 & 0.38            & 16 & 0 & 51 & 11 & 1.0 & 0.59            & 12 & 0 & 49 & 24 & 1.0 & 0.33            \\
\hline

\hline
\hline
\multicolumn{24}{||c||}{}\\
\multicolumn{24}{||c||}{{\bf When the agent is `unsure', this is counted as if it had determined the mutant to be equivalent}}\\
TP & FP & TN & FN & Prec. & Recall   & TP & FP & TN & FN & Prec. & Recall   & TP & FP & TN & FN & Prec. & Recall   & TP & FP & TN & FN & Prec. & Recall   \\
19 & 3 & 50 & 19 & 0.86 & 0.50           & 14 & 1 & 55 & 18 & 0.93 & 0.50           & 19 & 9 & 51 & 11 & 0.68 & 0.63           & 13 & 4 & 49 & 24 & 0.76 & 0.35           \\
\hline

\hline
\hline
\multicolumn{24}{||c||}{}\\
\multicolumn{24}{||c||}{{\bf When we include mutants that are syntactically identical in the numbers counted as determined to be equivalent}}\\
TP & FP & TN & FN & Prec. & Recall  &  TP & FP & TN & FN & Prec. & Recall   & TP & FP & TN & FN & Prec. & Recall   & TP & FP & TN & FN & Prec. & Recall   \\
41 & 3 & 50 & 19 & 0.93 & 0.58          &  42 & 1  & 55 & 18 & 0.98 & 0.70          & 47 & 9  & 51 & 11 & 0.84 & 0.81          & 31 & 4  & 49 & 24 & 0.86 & 0.56          \\
\hline

\hline
\hline
\multicolumn{24}{||c||}{}\\
\multicolumn{24}{||c||}{{\bf When we additionally include, as determined to be equivalent, those mutants that are syntactically identical after stripping any comments added by the generation agent}}\\
TP & FP & TN & FN & Prec. & Recall  & TP & FP & TN & FN & Prec. & Recall   & TP & FP & TN & FN & Prec. & Recall   & TP & FP & TN & FN & Prec. & Recall   \\
41 & 3 & 67 & 2  & 0.93 & 0.95          & 42 & 1 & 69 & 4  & 0.93 & 0.91           & 47 & 1 & 66 & 0  & 0.99 & 1.0            & 53 & 4  & 49 & 2  & 0.93      & 0.96      \\
\hline
\end{tabular}

\label{tab:equivalence_each}
\end{table*}

\section{The Importance of Mutation Testing}
\label{sec:why_mutation}
It is well known in the literature on testing, both theoretically and empirically,  
that mutation adequacy criteria outperform traditional structural criteria, such as line coverage and branch coverage \cite{mike:icse17,mpetal:mutation-advances}.
In this section, we present results that further underline these previous research findings with direct experience from the deployment of mutation testing on large scale industrial systems.

Table~\ref{tab:coverage_is_not_enough} reports line coverage results for the tests that kill non-equivalent mutants.
For Messenger, Instagram, Wearables and WhatsApp the `likely actual' number  is based on the ground truth equivalent proportion for these platforms in Table~\ref{tab:all_different_types_of_mutants_in_it_look_i_want_that_one_for_my_Christmas}. 
For the other platforms, it is based on the overall average  ground truth proportion over all four platforms.
The final three columns show the  results for TestGen-LLM \cite{mhetal:TestGen-LLM} generated tests, as proportions for comparison with the corresponding proportions for \toolname~generated tests. 
TestGen-LLM  does not target specific faults, but merely attempts to generate tests that will acquire extra coverage.

As Table~\ref{tab:coverage_is_not_enough} shows, a large proportion (49\%) 
of test cases that uniquely additionally kill a mutant, do not also add line coverage.
Clearly, these tests have value, since they catch faults that would otherwise go undetected.
However, if we were to judge test cases solely on the basis of the coverage they add, 
then such valuable test cases would be wrongly discarded.

The results in Table~\ref{tab:coverage_is_not_enough}  also reveal that, 
although TestGen-LLM generates tests for a higher proportion of classes compared to \toolname~(32\% vs. 5.3\%), it nevertheless, succeeds in killing a far smaller proportion of mutants (2.4\% vs 15\%).
This is because \toolname~ specifically targets the mutants, whereas TestGen-LLM does not.

From these results, we conclude that targeting mutants can also elevate coverage, but targeting coverage will be inadequate to kill  mutants.
Although this is already known from the research literature \cite{mpetal:mutation-advances,mike:icse17}, 
that literature is based on laboratory controlled experiments with non-industrial systems.
Our findings thus augment the existing literature and add weight to  claims 
for the importance of mutation  testing, based on industrial practice and experience.

It is also worth noting that 70\% of the mutants left unkilled by TestGen-LLM reside in classes that do not have {\em any} TestGen-LLM test, let alone one that kills the mutant. 
Combined with the high proportion (51\%) of \toolname~tests that {\em do} also raise coverage, we conclude that
Mutation-Guided LLM-based Test Generation has attractive side benefits on coverage.

Specifically, we may be able to adapt our \toolname~approach to target coverage using mutants: 
Suppose we generate many mutants, 
of all different kinds, 
and at all different levels of abstraction, 
for a given method under test. 
We can then use each as a prompt to an LLM.
In this way,  mutants play a role similar to Retrieval Augmented Generation (RAG).
Given our results and the fact that  RAG is known to improve LLM performance on other Software Engineering tasks~ \cite{mhetal:LLM-survey}, we believe mutation-as-RAG for coverage  is likely to prove attractive for coverage as well as for targeting specific faults.


\begin{table*}
\caption{The numbers of tests that add coverage as well as detecting mutated faults. 
This gives an upper bound on the number of faults we might miss by merely targeting new coverage alone. 
Approximately half of all tests generated, although finding faults missed by all existing tests, do not add line coverage. 
Had we used coverage as our sole test adequacy criterion,  the platforms would thus have continued to be vulnerable to regressions denoted by up to approximately half of the faults.
}
\vspace{-4mm}
  \centering
  \begin{tabular}{|l|c|r|r|c|r|r|r|r|}
    \hline
    Platform          & Number of      & Number of  & {\%}age     & {\%}age         & Number     & {\%}age   & {\%}age        & {\%}age   \\
                     & (believed;      & mutant-    & of classes  & mutant          & ({\%}age)  & Classes   & TestGen-       & unkilled  \\
                     & likely actual)  & killing    & with a      & kill rate       & of tests   & with a    & LLM kill       & mutants   \\ 
                     & non-equivalent  & tests      & mutant-     & on (believed;   & that do    & TestGen-  & rate on        & with no   \\
                     &   mutants       &            & killing     & likely actual)  & not add    & LLM test  & (believed;     & TestGen-  \\
                     &                 &            & test        &                 & coverage   &           & likely actual) & LLM test  \\
    \hline
    Facebook Feed    &    164;   133   &  15        & 4.3\%       &  9\%; 11\%    &   9 (60\%)   &   33\%    & 3.0\%; 3.7\%   &   71\%    \\
    Messenger        &  1,395; 1,155   & 196        & 5.9\%       & 14\%; 17\%    &  84 (43\%)   &   31\%    & 2.0\%; 2.4\%   &   71\%    \\
    Instagram        &    742;   687   & 122        & 7.2\%       & 16\%; 18\%    &  77 (55\%)   &   27\%    & 2.9\%; 3.1\%   &   74\%    \\
    Aloha            &     90;    74   &   9        & 5.1\%       & 10\%; 12\%    &   4 (44\%)   &   27\%    & 1.0\%; 1.2\%   &   76\%    \\
    Wearables        &  1,314; 1,007   &  45        & 1.6\%       &  3\%;  4\%    &  32 (71\%)   &   31\%    & 1.9\%; 2.5\%   &   70\%    \\
    Cross-app        &    270;   275   &  35        & 4.3\%       & 13\%; 13\%    &  20 (57\%)   &   28\%    & 2.8\% 2.7\%    &   75\%    \\
    Oculus           &    150;   121   &  14        & 4.3\%       &  9\%; 12\%    &   3 (21\%)   &   35\%    & 4.9\%; 6.1\%   &   70\%    \\

    WhatsApp         &    535;   445   & 135        & 10.6\%      & 25\%; 30\%    &  48 (36\%)   &   44\%    & 3.5\%; 4.2\%   &   55\%    \\
    \hline
    Totals           &  4,660; 3,897   & 571        & 5.3\%       & 12\%, 15\%    & 277 (49\%)   &   32\%    & 2.0\%; 2.4\%   &   70\%    \\
    \hline
  \end{tabular}

\label{tab:coverage_is_not_enough}
\end{table*}

\section{Related Work}
Given the strong empirical evidence  for the  predictability of code   \cite{hindle:naturalness,ebetal:psh-fse14,gabel:uniqueness-tiny},
it is unsurprising that predictive language models have proved effective at generating usable code.
As a result, LLMs now play a code-generation role in a significant number of applications across the spectrum of software engineering activity \cite{mhetal:LLM-survey}. 

Software  engineers' acceptance rates for LLM-generated code have been widely studied.
For example,  researchers at Google focussed on factors that may influence trust in AI-powered code completions
\cite{brown:identifying}, while it was recently reported that Copilot had an acceptance rate of approximately 30\%~\cite{copilot_acceptance_rate_june_2023}.

However, such studies concern the code completion use case,  which does not come with any assurances (the code completions may not even compile).
By contrast, \toolname~uses Assured LLM-Based Software Engineering (Assured LLMSE) \cite{mhetal:intense24-keynote}.
That is, \toolname~ provides assurances about the semantics and usefulness of the tests it proposes. 
Its proposals are also {\em whole} compilable code units (not merely completion fragments), and
it is deployed on-demand (to generate tests targeting classes of faults of particular interest), 
rather than opportunistically (to suggest code completions).
The fact that \toolname~provides these assurances, and its different deployment route, mean that we can expect a higher overall acceptance rate from \toolname, than we would for code completion technologies.

While code completions are deployed directly into the IDE in real time (online), 
\toolname~ is an offline technology that proposes code updates 
in exactly the same way that engineers would  propose them: 
through Continuous Integration (CI) and standard code review.
This has the advantage that the decision to accept machine-generated code 
has an audit trail, 
which is missing for LLM-based code completions.
By contrast, 
LLM-based code completion suggestions accepted by engineers are attributed to the engineer by the CI audit trail; details of the LLM's role in constructing production code through completion suggestions  is thus unavailable. 

There is already a considerable body of literature on LLMSE \cite{mhetal:LLM-survey}.
The subset of this literature devoted specifically to software testing is too large to survey in detail here.
Instead, we refer readers to recent surveys~\cite{wang:llm-test-survey,mhetal:LLM-survey}.
In this section, we focus specifically on the application of LLMs to mutation testing.

Traditionally~\cite{yjmh:analysis}, mutation testing techniques have used 
sets of pre-defined rule-based operators to construct mutants.
However, it has been widely observed ~\cite{richter2022learning,mjetal:are-mutants} 
that such pre-defined rule-based mutation operators are ill-suited to the task of generating {\em realistic} faults.
This has led researchers to consider ways to make more realistic mutants \cite{yjmh:homt}, a problem to which language models have recently been applied \cite{tip2024llmorpheus,wang2024exploratory}.
\toolname~ represents an extension of this direction that additionally seeks to generate a  set of faults, relevant to a chosen issue of concern.

Mutation testing has also been used as a way to evaluate language models themselves~\cite{li2024mutation} and to mutate compiler inputs to  test compilers \cite{italiano:finding}.
In this work, the desirable property of mutants is the way that they generate near neighborhoods of programs that are similar to the program that they mutate, both syntactically and semantically. 
This allows researchers to explore properties of language models when applied to code, such as their ability to `understand' the code.
This `near neighborhood' property has also been used in combination with language models~\cite{brownlee2023enhancing} to generate mutants to help improve other software engineering technologies, such as Genetic Improvement~\cite{Petke:gisurvey}.

The equivalent mutant problem has been widely studied  \cite{yjmh:analysis,madeyski2013overcoming}, and we are not the first to consider 
using language models to tackle it.
Tian et al. \cite{tian2024large}  recently studied the use of language models to detect equivalent mutants for traditional rule-based mutation testing operators on MutantBench. 
MutantBench~\cite{van2021mutantbench} is a set of mutant-original-code pairs, written in C/C++ and Java, introduced in 2021 by van Hijfte and Oprescu. 
Tian et al. found that LLM-based equivalence detection outperformed traditional non-LLM-based equivalence detection by 35\% for F1 score (the harmonic mean of precision and recall). In particular, they highlighted the high-performance of fine-tuned embeddings.

All the previous work on LLMs for mutation testing has concerned benchmark problems and empirical analysis under laboratory conditions.
The present paper is the first to report the deployment of LLM-based mutation testing at scale in industry.
Non-LLM-based mutation testing has previously been evaluated in industrial application settings, 
at  Meta~\cite{beller:what} 
and at Google ~\cite{petrovic2021practical,petrovic2018state,petrovic2018industrial}.
However, in both these previous industrial deployments, there was no attempt to automatically generate tests to 
{\em kill} the mutants.
The challenging task of constructing tests to kill mutants
was left to engineers, who also therefore had to contend with the equivalent mutant problem.
Unlike these previous industrial  deployments, \toolname~is also the first to automatically generate tests to kill mutants; a problem which has been studied in the research literature~\cite{carlos:test,mhetal:shom}, but not previously deployed at 
scale in industry.

The work on \toolname~reported here, follows a succession of automated analysis and testing work  deployed at Meta.
Our initial focus was end-to-end test generation tools such as Sapienz \cite{mhetal:ssbse18-keynote,mao:sapienz:16}, 
and static analysis tools such as Infer \cite{movefast}, followed by
simulation-based test generation for testing interacting user communities \cite{jaetal:mia,tuli:simulation,jaetal:ease21-keynote}.
While much of this previous work has been concerned with system level testing, 
we have also worked on unit level test generation techniques using observations \cite{mhetal:TestGen-obs} and language models \cite{mhetal:TestGen-LLM}.
However, this previous unit test generation work was based on the sole objective of increasing coverage, which meant it could not target specific classes of faults as \toolname~does.

\section{Open Problems}
\label{sec:future}
\label{sec:open_probelms_for_research}
We outline directions for future work, hoping to stimulate the 
research community with new open research questions.

\noindent
{\bf Mutant Equivalence}: we found that the  \PrivacyMutants~generated by language models tend to be bimodal, with the consequence that equivalent mutants are surprisingly easy to detect (See Section~\ref{sec:equivalance}). 
It is possible that these results are merely specific 
to Kotlin,  
to Android, 
to Meta, 
or to \PrivacyMutants. 

However, there is nothing in our approach that suggests that the results would fail to generalize. 
If they {\em were} to generalize, this would be an  important finding for the mutation testing research agenda, 
given the significant impediment to deployment  hitherto posed by mutant equivalence.

\noindent
{\bf Mutant Relevance}: We have been  able to generate specific mutants that capture similar faulty behaviors as those previously witnessed. 
However, in some cases, anecdotally, looking at the faults generated, 
it seemed that the mutant was related to the general class of fault simulated, 
but was not an example of the specific instance. 
One difficulty here is that we have no way to consistently and reliably measure problem similarity or relevance.

More research is needed to define what it means for one fault to be similar to another.
We also need approaches that use such similarity metrics to guide  agenetic LLM  workflows, e.g.,  with fine-tuning, 
prompt engineering, re-prompting, and/or Chain-of-Thought, to ensure that the mutants generated are relevant to the original fault.

\noindent
{\bf Detecting existing faults}: our approach hardens against future regressions.
This means that the current version of the system under the test is used as the regression oracle \cite{alshahwan:software}: 
a test is deemed to pass  after some change if the test behaves the same before and after the change.
This allows us to protect against future regressions, but it cannot detect {\em existing} faults residing in the code base.
To do this requires us to tackle the well known Oracle problem \cite{ebetal:oracle}.

It would be very exciting if an approach can be found to infer oracles for Targeted LLM-based Mutation-Guided Test Generation.
Such an advance would be highly impactful because it would allow us to search for existing classes of faults.
Although there has been much previous work on oracle inference ~\cite{tuli:simulation,ibrahimzada2022perfect,oliveira:automated,zhang:search-based-MR}, we need higher precision to avoid wasting engineers' time on false positives.
Generating faults/tests for specific issues of concern, as \toolname~does, may help oracle generation.

\section{Conclusions}
\label{sec:conclusions}
Mutation testing has been the subject of research for  five decades \cite{howden:theory,demillo78,yjmh:analysis,mpetal:mutation-advances}, so
 has clearly retained intellectual appeal and researchers' curiosity.
Despite this, it has proved challenging to deploy mutation testing in industry \cite{beller:what,alshahwan:software,mhetal:issta19}.

Traditionally, mutants have been constructed using simple {\em rule-based} approaches in order to  {\em assess} test suites, largely {\em written by humans}. 
However, software engineers need automatically generated tests, that are  relevant to a specific pressing concern; their time would be better spent articulating these pressing concerns, rather than trying to construct test cases that should, instead, be generated by a machine.
Fortunately, two crucial recent advances, drawn together in \toolname, make this possible:
Automated test generation to kill generated mutants, coupled with language models' ability to generate highly-relevant mutants.
In particular, we found that LLMs help us to overcome existing barriers~\cite{beller:what} to deployment of mutation testing. Specifically, they

\begin{enumerate}
    \item Allow us to generate highly realistic faults, using their ability to convert text (the concern or issue) into code (simulated faults from which we generate tests);
    \item Provide an additional agent to weed out equivalent mutants, which is especially powerful when combined with simple static analyses as a pre-processing step;
    \item Provide a way to automatically generate unit tests to kill the mutants.
\end{enumerate}

This paper presented results from deployment of \toolname~ at Meta. 
Neither LLM-based test generation, nor LLM-based mutant generation is new, 
but this paper is the first to report on their combined deployment on large scale industrial systems.
We believe this form of automated test generation is highly aligned with modern software development and deployment.
It supports software engineers who must contend with many competing and conflicting concerns, often expressed in natural language in vague, incomplete and even contradictory ways.
Our results also suggest Mutation-as-RAG will  prove impactful in optimizing  for structural coverage criteria. 


\section*{ACKNOWLEDGMENTS}
We  would like to thank the Meta's Llama team and leadership of the Fundamental Artificial Intelligence Research (FAIR), Developer Infrastructure (DevInfra) and Product Compliance and Privacy  teams for supporting this work and the Meta Software Engineers who so freely and kindly gave of their time, experience, and  expertise in reviewing the tests automatically generated by \toolname.

\balance
%
\bibliographystyle{ACM-Reference-Format}
\bibliography{slice}

\end{document}